# Anomalous Hall Effect in three ferromagnets: $EuFe_4Sb_{12}$, $Yb_{14}MnSb_{11}$, and $Eu_8Ga_{16}Ge_{30}$

Brian C. Sales, Rongying Jin, and David Mandrus

Condensed Matter Sciences Division, Oak Ridge National Laboratory, Oak Ridge, Tennessee 37831

Peter Khalifah

Condensed Matter Sciences Division, Oak Ridge National Laboratory, Oak Ridge, Tennessee 37831 and

Department of Chemistry, University of Massachusetts, Amherst, Massachusetts 01003

The Hall resistivity ($\rho_{xy}$), resistivity ($\rho_{xx}$), and magnetization of three metallic ferromagnets are investigated as a function of magnetic field and temperature. The three ferromagnets, $EuFe_4Sb_{12}$ ($T_c \approx 84$ K), $Yb_{14}MnSb_{11}$ ($T_c \approx 53$ K), and $Eu_8Ga_{16}Ge_{30}$ ($T_c \approx 36$ K) are Zintl compounds with carrier concentrations between 1 x $10^{21}$ cm$^{-3}$ and 3.5 x $10^{21}$ cm$^{-3}$. The relative decrease in $\rho_{xx}$ below $T_c$ [$\rho_{xx}(T_c)/\rho_{xx}(2\text{ K})$] is 28, 6.5, and 1.3 for $EuFe_4Sb_{12}$, $Yb_{14}MnSb_{11}$, and $Eu_8Ga_{16}Ge_{30}$ respectively. The low carrier concentrations coupled with low magnetic anisotropies allow a relatively clean separation between the anomalous ($\rho'_{xy}$), and normal contributions to the measured Hall resistivity. For each compound the anomalous contribution in the zero field limit is fit to $a\rho_{xx} + \sigma_{xy}\,\rho_{xx}^2$ for temperatures $T < T_c$. The anomalous Hall conductivity, $\sigma_{xy}$, is -220 ± 5 ($\Omega^{-1}$ cm$^{-1}$), -14.7 ± 1 ($\Omega^{-1}$ cm$^{-1}$), and 28 ± 3 ($\Omega^{-1}$ cm$^{-1}$) for $EuFe_4Sb_{12}$, $Yb_{14}MnSb_{11}$, and $Eu_8Ga_{16}Ge_{30}$ respectively and is independent of temperature for $T < T_c$ if the change in spontaneous magnetization (order parameter) with temperature is taken into account. These data are consistent with recent theories[1,2,3,45] of the anomalous Hall effect that suggest that even for stochiometric ferromagnetic crystals, such as those studied in this article, the intrinsic Hall conductivity is finite at T = 0, and is a ground state property that can be calculated from the electronic structure.

75.47.-m, 75.50.Cc

## Introduction

The origin of the large "anomalous" contribution to the Hall resistivity of ferromagnets has remained a source of confusion since the seminal experiments of Hall[6,7] in the 1880's. The quantity measured in a standard DC Hall experiment is the Hall resistivity $\rho_{xy} = E_y/J_x = V_y d/I_x$ where a DC current $I_x$ flows through a rectangular slab of thickness d in the x direction, the voltage is measured across the sample in the y direction with a magnetic field B applied along the z direction. In materials with no significant magnetism, the Hall resistivity is proportional to the applied magnetic field and describes the effect of the Lorentz force on the motion of the free carriers. In simple one band materials the Hall resistivity can be used to measure the type (electrons or holes) and number of carriers. In ferromagnetic compounds or materials with a substantial magnetic susceptibility, there is an additional contribution to the Hall resistivity that is proportional to the magnetization, M, of the material and is usually at least one order of magnitude larger than the Lorentz term. For a magnetic material the Hall resistivity is described by $\rho_{xy} = R_oB + R_s4\pi M = R_oB + \rho_{xy}'$ where $R_oB$ is the ordinary contribution and $\rho_{xy}'$ describes the anomalous contribution to the Hall effect (AHE). In a good metal (like iron) the coefficient $R_s$ is much larger than $R_o$. Various theories[8,9,10,11] of the origin $R_s$ have shown that $R_s$ can be described by intrinsic and extrinsic contributions that are proportional to $\rho^2$ or $\rho$, respectively, where $\rho$ is the zero field resistivity $\rho = \rho_{xx}(0)$. The extrinsic contribution[8,9] accounts for the skew scattering of carriers by magnetic impurities and defects but is hard to quantitatively model in real materials. The pioneering work of Karplus and Luttinger[11], however, showed there was an intrinsic contribution to the AHE that arose from the spin-orbit coupling of Bloch bands, i.e. a contribution that in principle could be extracted from the calculated electronic structure. Thus the anomalous Hall resistivity *in zero applied magnetic field*, $\rho_{xy}'$ can be parameterized[12] by $\rho_{xy}' = \sigma_{xy}' \rho^2 + a\rho$, where $\sigma_{xy}'$ is the intrinsic anomalous Hall conductivity, and *a* describes the extrinsic contribution of skew scattering. We emphasize that in zero applied field the normal Hall contribution is zero and the anomalous contribution vanishes when $T > T_c$. To be consistent with the notation used in the theoretical calculations we set $\sigma_{xy}' = \sigma_{xy}$. In the case of Fe metal, the calculations of Yao et al. [2] predict an intrinsic Hall conductivity of $\sigma_{xy} = 751\ (\Omega\text{cm})^{-1}$ at T=0 and $\sigma_{xy} = 734\ (\Omega\text{cm})^{-1}$ at T = 300 K as compared to the room temperature experimental[13] value of 1032 $(\Omega\text{cm})^{-1}$. An interesting consequence of this calculation is that temperature has very little effect on $\sigma_{xy}$ as long as there is no significant change in the magnetization over the temperature range of interest. This observation has important consequences on the best way to separate the intrinsic and extrinsic contributions from experimental data.[12,14]

In the present article, we investigate the intrinsic and extrinsic contributions to the AHE in three unusual ferromagnets: $EuFe_4Sb_{12}$, $Yb_{14}MnSb_{11}$, and $Eu_8Ga_{16}Ge_{30}$. All three ferromagnets are stochiometric Zintl compounds[15,16] which implies both ionic and covalent bonds. The compounds can be regarded as heavily doped semiconductors or bad metals with typical carrier concentrations of 1 to 3.5 x $10^{21}$ carriers/$cm^3$. The dominant carriers are holes in $EuFe_4Sb_{12}$ and $Yb_{14}MnSb_{11}$ and electrons in $Eu_8Ga_{16}Ge_{30}$. Although the three compounds are magnetically soft ferromagnets, the crystal structures are relatively complex with 34, 208, and 54 atoms in the conventional unit cell, respectively. One of the goals of the present work is to see how well current thinking about the origin of the AHE applies to more complex materials.

## Synthesis and Experimental Methods

The synthesis of $EuFe_4Sb_{12}$, $Yb_{14}MnSb_{11}$ and $Eu_8Ga_{16}Ge_{30}$ has been described in detail previously.[17,18,19,20] Briefly, $EuFe_4Sb_{12}$ was prepared directly from the elements in a carbon-coated, evacuated and sealed silica tube. The tube was heated to 1030 °C for 40 h, quenched into a water bath, and then heated at 700 °C for one week to form the correct cubic skutterudite phase. The resulting powder was ball milled in Ar gas, and then hot pressed in vacuum in a graphite die into a dense polycrystalline solid. X-ray diffraction confirmed that the samples were single phase with the cubic skutterudite structure (space group Im-3, lattice constant $a$ = 0.917 nm, 34 atoms per unit cell). The estimated filling of the Eu site was 0.95, a value comparable to that found in single crystals.[21,22] Heat capacity, magnetization and resistivity measurements on the hot pressed sample indicate $EuFe_4Sb_{12}$ becomes ferromagnetic below 84 K. Hall data from $EuFe_4Sb_{12}$ are obtained on a thin polycrystalline rectangular plate.

Crystals of $Yb_{14}MnSb_{11}$ were grown from a Sn flux using the method described by Fisher et al.[18] This phase, originally reported by Chan et al.[23] crystallizes with a tetragonal lattice in the space group I41/*acd* with $a$ = 1.661 nm and $c$= 2.195 nm, and 208 atoms in the conventional unit cell. Heat capacity, resistivity, and magnetization data on $Yb_{14}MnSb_{11}$ crystals indicate ferromagnetism below 53 K.[18,19] Hall data from $Yb_{14}MnSb_{11}$ are taken on a thinned single crystal plate with H // (110) and the current along the **c** direction.

The cubic clathrate compound $Eu_8Ga_{16}Ge_{30}$ (space group Pm-3n, $a$ = 1.070 nm, 54 atoms in unit cell) could be prepared by several methods, the simplest of which was the direct arc melting together of the elements on a water-cooled copper hearth in an Ar atmosphere. Crystals are also grown by cooling a stochiometric melt of the elements in a carbon coated silica ampoule.[20] A small unoriented single crystal plate ( H ≈ // (4 3 2)) of $Eu_8Ga_{16}Ge_{30}$ is used in the present experiments with $T_c \approx 36$ K.

Hall effect, resistivity, and heat capacity data are taken using a Physical Property Measurement System (PPMS) from Quantum Design. Hall and resistivity data are obtained using a standard six lead method and either rotating the sample by 180° in a fixed magnetic field or by sweeping the direction of the field from positive to negative values. Low resistance electrical contacts are made to $EuFe_4Sb_{12}$ with silver epoxy, and to $Yb_{14}MnSb_{11}$ with silver paste. For the $Eu_8Ga_{16}Ge_{30}$ crystals, however, the surface is first etched with Ar ions and then sputtered coated with gold pads. Silver paste is used to attach 0.025 mm diameter Pt wires to the gold pads. The Hall data from each material is qualitatively the same regardless of the orientation of the crystal with respect to the current or field directions. Magnetization data are taken using a commercial SQUID magnetometer from Quantum Design.

## Magnetic Data and Background Information

Magnetization curves at 5 K for the three compounds are shown in Fig. 1. For each ferromagnet there is very little hysteresis and the full saturation moment is reached at relatively low applied magnetic fields (H < 1 T). The magnetic anisotropy is small (a few hundred gauss) for the cubic compounds ($EuFe_4Sb_{12}$ and $Eu_8Ga_{16}Ge_{30}$) while for $Yb_{14}MnSb_{11}$ the maximum anisotropy corresponds to a field of about 1 T (i.e. if the field is applied along the magnetically hard (110) direction it takes about 1 T to reach full saturation at 2 K). The magnetization data for $Yb_{14}MnSb_{11}$ shown in Fig 1 is taken with the applied field along the easy (001) direction.

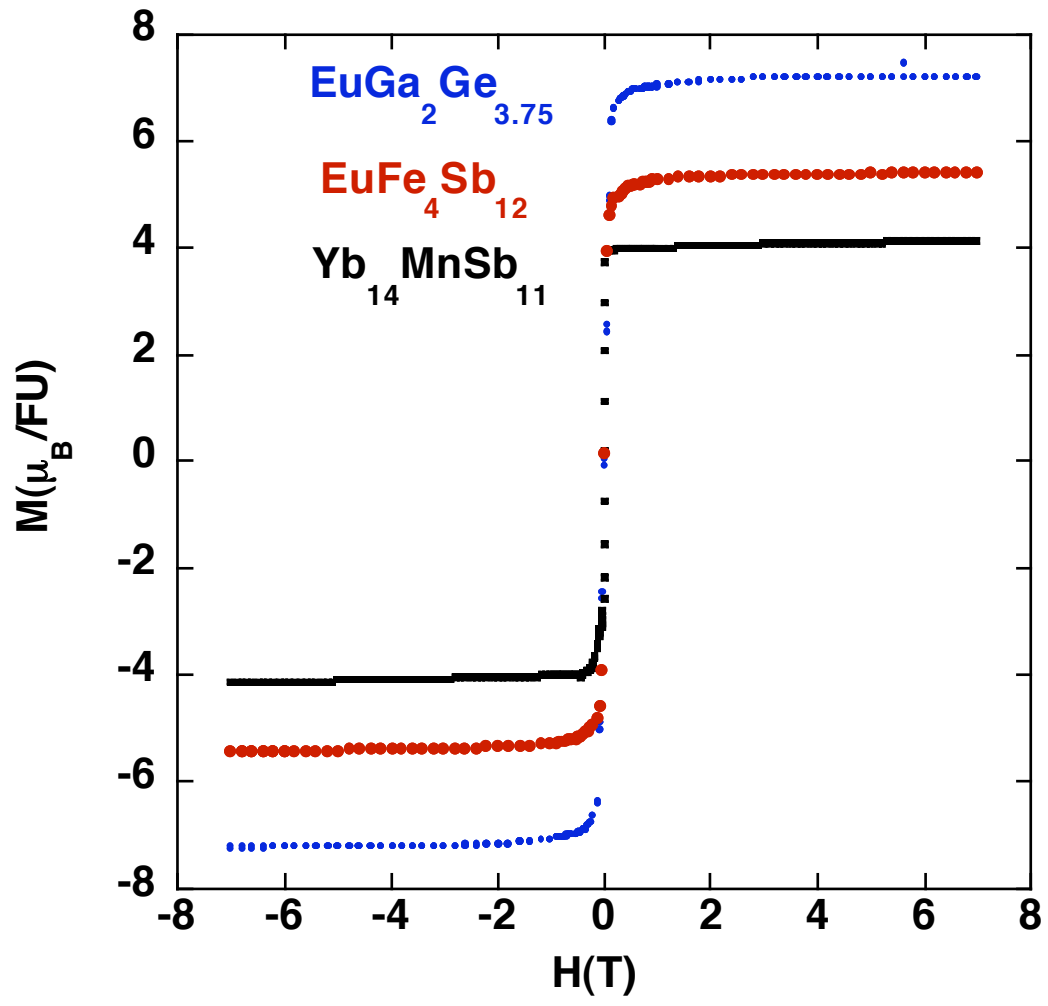

FIG. 1. Magnetization vs. magnetic field at 5 K for the three ferromagnetic compounds

The magnetization curve for $Eu_8Ga_{16}Ge_{30}$ is the simplest to understand since the saturation magnetization nearly corresponds to the ionic Hund's rule value of 7 $\mu_B$ per $Eu^{2+}$. The large separation distance between neighboring $Eu^{2+}$ ions in the $Eu_8Ga_{16}Ge_{30}$ structure (about 5.4 Å), and the poor bonding between the $Eu^{2+}$ ions and the Ga and Ge atoms forming a cage around each Eu suggests that the magnetic order in this material occurs via a Rudermann-Kittel-Kasuya-Yoshida (RKKY) indirect exchange interaction.[20,24,25] There are two types of Ge-Ga cages in the Type-I clathrate structure corresponding to polyhedra consisting of either 20 (small cage) or 24 (large cage) atoms of Ge or Ga with Eu ions residing near the center of each cage. In the large cage, the Eu is poorly bonded and moves off center to one of four nearly equivalent sites[20]. Remarkably, Mossbauer and RF absorption measurements show that the Eu ions actually tunnel among the four sites at a slow frequency of about 450 MHz![26] Elastic constant measurements and theory also suggest (although indirectly) a similar tunneling frequency.[27] The slow dynamics due to movement of the Eu ions in the large cages does not appear to affect the DC magnetic properties, but may contribute to the unusually low carrier mobility well below $T_c$.

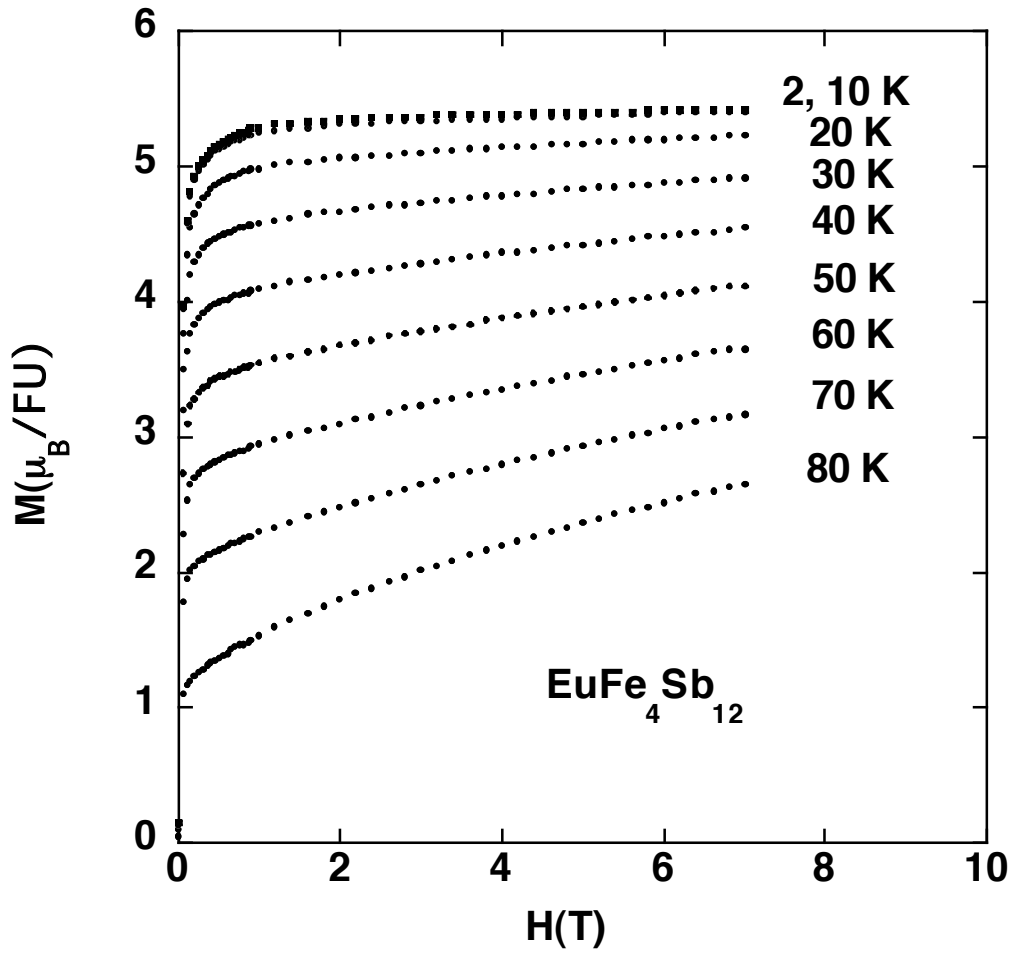

FIG. 2. Magnetization vs. magnetic field at various temperatures for the ferromagnet $EuFe_4Sb_{12}$.

The compound $Yb_{14}MnSb_{11}$ may be a rare example of an under screened Kondo lattice[19,28]. In the material all of the Yb ions are divalent and hence nonmagnetic. The magnetism comes from the Mn 3d electrons and the antiferromagnetic coupling of these electrons to holes in the Sb 5p bands. The shortest Mn-Mn distance in $Yb_{14}MnSb_{11}$ is 1.0 nm. The electronic state of the Mn is believed to be the $d^5$+hole configuration that is found, for example, in GaAs doped with Mn.[29] The antiferromagnetic coupling between a local magnetic moment and extended Bloch states can result in ferromagnetism due to the RKKY interaction, but can also give rise to Kondo physics. In $Yb_{14}MnSb_{11}$, the ferromagnetic ground state is accompanied by a substantial ( $\approx 20\ m_e$ ) mass enhancement of the holes near the Fermi energy. A reduction of the hole concentration in $Yb_{14}MnSb_{11}$ by chemical replacement of 5% of the Yb by La, lowers $T_c$ to 39 K, but increases the ferromagnetic saturation moment because there are fewer carriers to screen the Mn $d^5$ moment.[19]

The ferromagnetism of the filled skutterudite $EuFe_4Sb_{12}$ is more complex than we initially thought. The skutterudite compounds with divalent cations, such as $CaFe_4Sb_{12}$, are incipient band ferromagnets[30,31] in the sense of the Stoner model. A further increase in the carrier concentration by the replacement of divalent Ca by monovalent Na, results in a ferromagnet[32] with a $T_c$ of 85 K, surprisingly close to the 84 K of $EuFe_4Sb_{12}$. This suggests that the polarization of the Fe 3d bands caused by the substitution of Ca by magnetic Eu ions is enough to drive the compound ferromagnetic. In addition, careful Mossbauer[26] and x-ray magnetic circular dichromism measurements[33] on our $Eu_{0.95}Fe_4Sb_{12}$ samples indicate that about 10-15% of the Eu is $Eu^{3+}$, which has a nonmagnetic J=0 groundstate. The dichromism measurements also show a small moment of $\approx$0.1- 0.2 $\mu_B$ per Fe, and that this moment is oriented opposite to the $Eu^{2+}$moment. Taken together, all of these effects predict a saturation moment of about 5 $\mu_B$ per formula unit of $EuFe_4Sb_{12}$ , which is near the value measured in Fig 1.

As discussed in the introduction, the Hall resistivity of a ferromagnet is described by $\rho_{xy} = R_oB + R_s4\pi M$, where the second term is the anomalous contribution to the Hall resistivity. If a relatively small magnetic field is applied (H< 1T) to a soft ferromagnet, essentially all of the domains will be aligned and the Hall resistivity will be dominated by the second term. As the temperature is increased to an appreciable fraction of $T_c$, however, the spontaneous value of M will depend on how the order parameter (spontaneous magnetization within a single domain) changes with temperature. This behavior is illustrated by the magnetization curves shown in Fig 2 for $EuFe_4Sb_{12}$. For temperatures not too close to $T_c$ ($T < 0.9\ T_c$ ) the approximate variation of the order parameter with T can be estimated by extrapolating the high field portion of the magnetization curves back to H =0. Alternately, the magnetization versus temperature can be measured at the lowest field (H ≈ 1 T) where all of the domains are aligned. Both analyses give essentially the same variation of $M/M_o$ with $T/T_c$ as shown in Fig. 3.

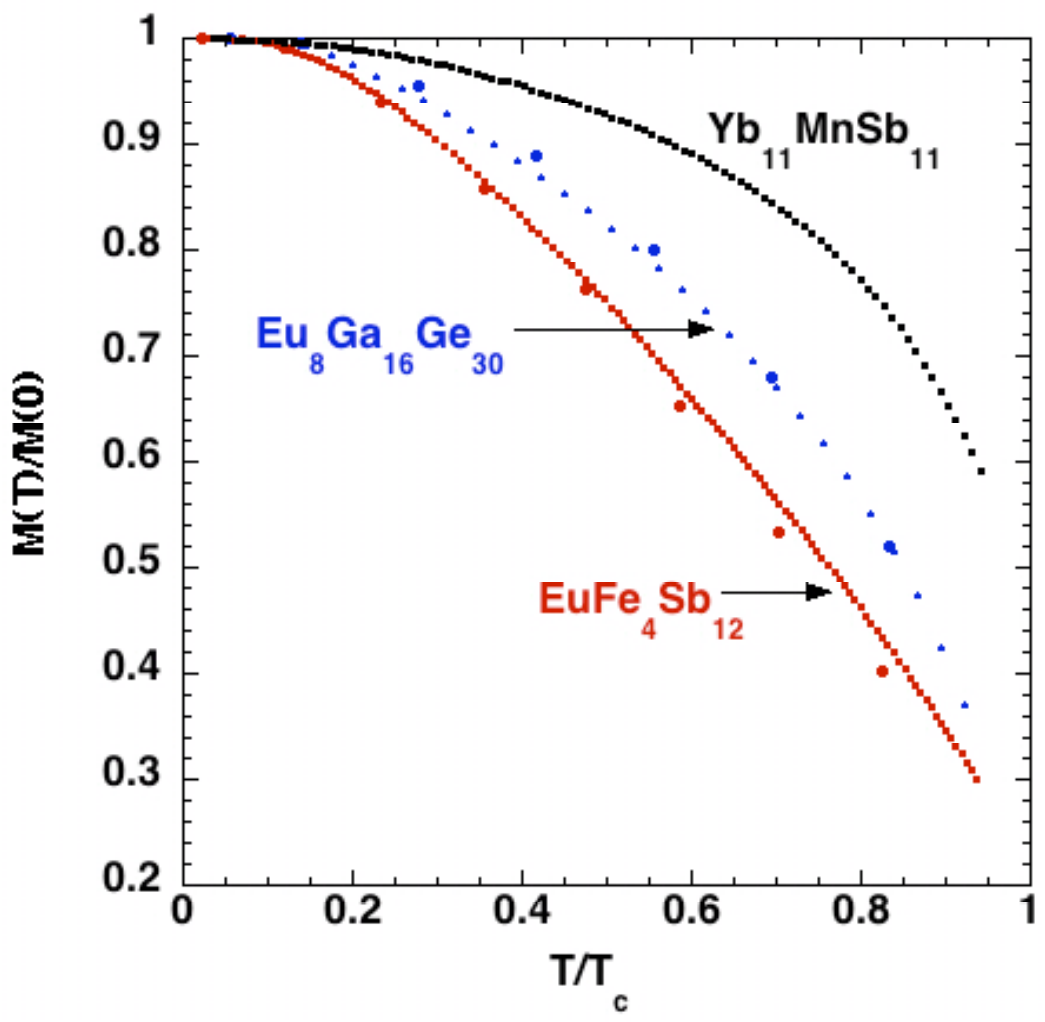

FIG. 3. Normalized magnetization M(T)/M(0) vs. $T/T_c$ for the three ferromagnets. The larger points correspond to estimates of M(T)/M(0) obtained from magnetization curves such as shown in Fig. 2. The remaining data were obtained from magnetization versus temperature measurements with an applied field of 1 Tesla.

## Resistivity and Hall Data

The dc resistivity versus temperature for each ferromagnet is shown in Fig 4a. At room temperature the magnitude of the resistivity is similar for all three compounds with values of 0.38 mΩ cm, 0.54 mΩ cm , and 1.7 mΩ cm for $EuFe_4Sb_{12}$, $Eu_8Ga_{16}Ge_{30}$ and $Yb_{14}MnSb_{11}$, respectively. The resistivity of tetragonal $Yb_{14}MnSb_{11}$ is measured with the current along the (001) direction. These values are suggestive of a heavily doped semiconductor or a bad metal. Below the ferromagnetic transition temperature of each compound there is a rapid decrease in the resistivity due to the loss of spin disorder scattering. The relative decrease in the resistivity below $T_c$ is defined as $[\rho(T_c)/\rho(2\ K)]$ and is 28, 6.5, and 1.3 for $EuFe_4Sb_{12}$, $Yb_{14}MnSb_{11}$, and $Eu_8Ga_{16}Ge_{30}$ respectively. The relatively small change in the resistivity with temperature for the $Eu_8Ga_{16}Ge_{30}$ single crystal is probably due to the random distribution of Ga and Ge on the framework sites and the disorder due to the Eu atoms in the larger cages slowly tunneling or hopping among 4 off center positions. The normalized resistivity for each compound below $T_c$ is shown in Fig 4b. These data illustrate the relatively large decrease in the resistivity of

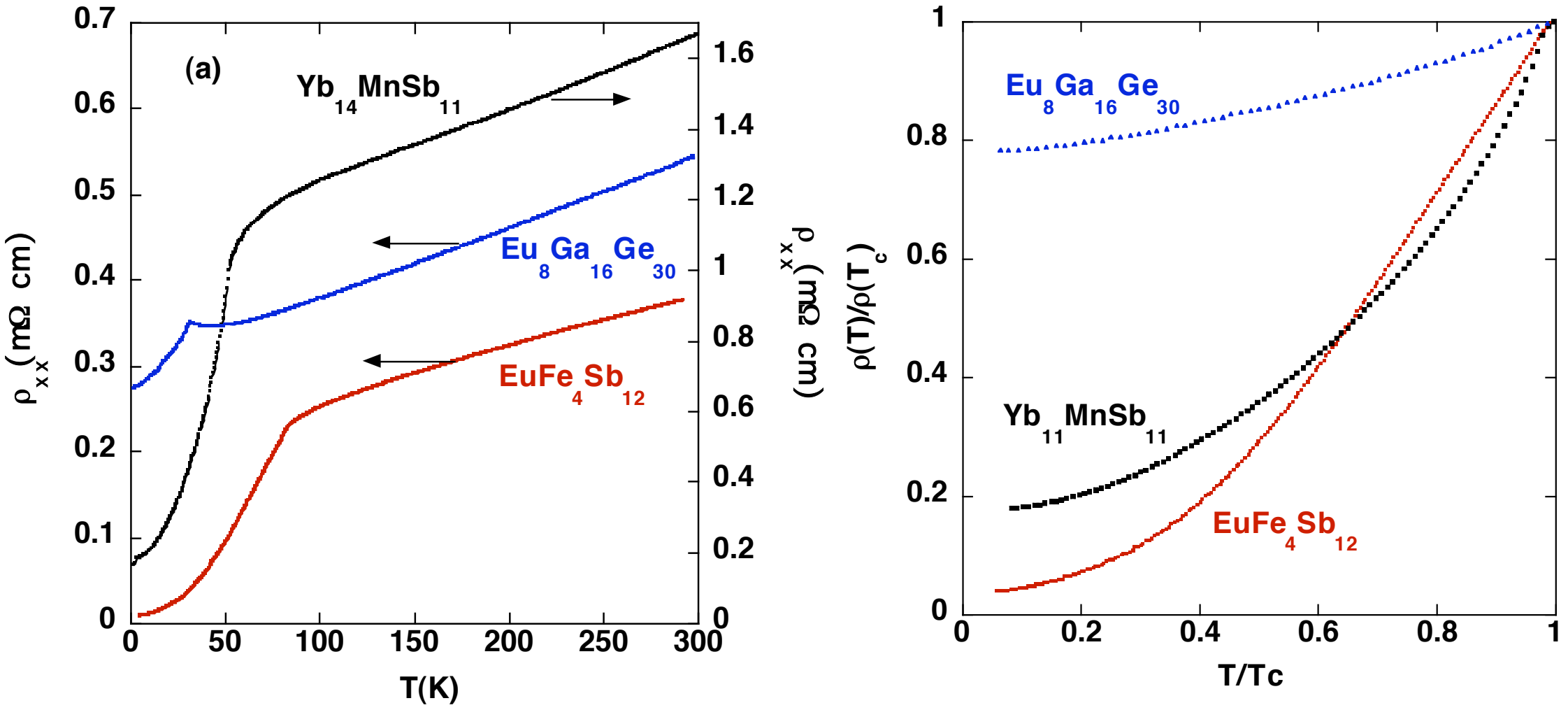

$EuFe_4Sb_{12}$ and $Yb_{14}MnSb_{11}$ below $T_c$, which makes possible an accurate evaluation of the dependence of the AHE on $\rho$ and $\rho^2$.

FIG. 4. (a) Resistivity vs. temperature for the three ferromagnets. The data for the $Yb_{14}MnSb_{11}$ crystal is taken with the current along the *c* axis. (b) Normalized resistivity vs. $T/T_c$ for the three ferromagnets

The Hall resistivity at 7 Tesla versus temperature is shown in Fig. 5 for all three compounds. For $EuFe_4Sb_{12}$ and $Eu_8Ga_{16}Ge_{30}$ the Hall resistivity is relatively constant for temperatures from 150 to 300 K with values of $2 \times 10^{-6}$ $\Omega$ cm and $-2.5 \times 10^{-6}$ $\Omega$ cm respectively. Assuming a single carrier band yields a carrier concentration of $2.2 \times 10^{21}$ holes $cm^{-3}$ for $EuFe_4Sb_{12}$ and $1.75 \times 10^{21}$ electrons $cm^{-3}$ for $Eu_8Ga_{16}Ge_{30}$. The decrease in the Hall resistivity of $EuFe_4Sb_{12}$ below 40 K is related to a large anomaly in the elastic constants[34], which may reflect a subtle change in the crystal structure. Only the Hall resistivity of $Yb_{14}MnSb_{11}$ well above $T_c \approx 53$ K has a significant contribution from the

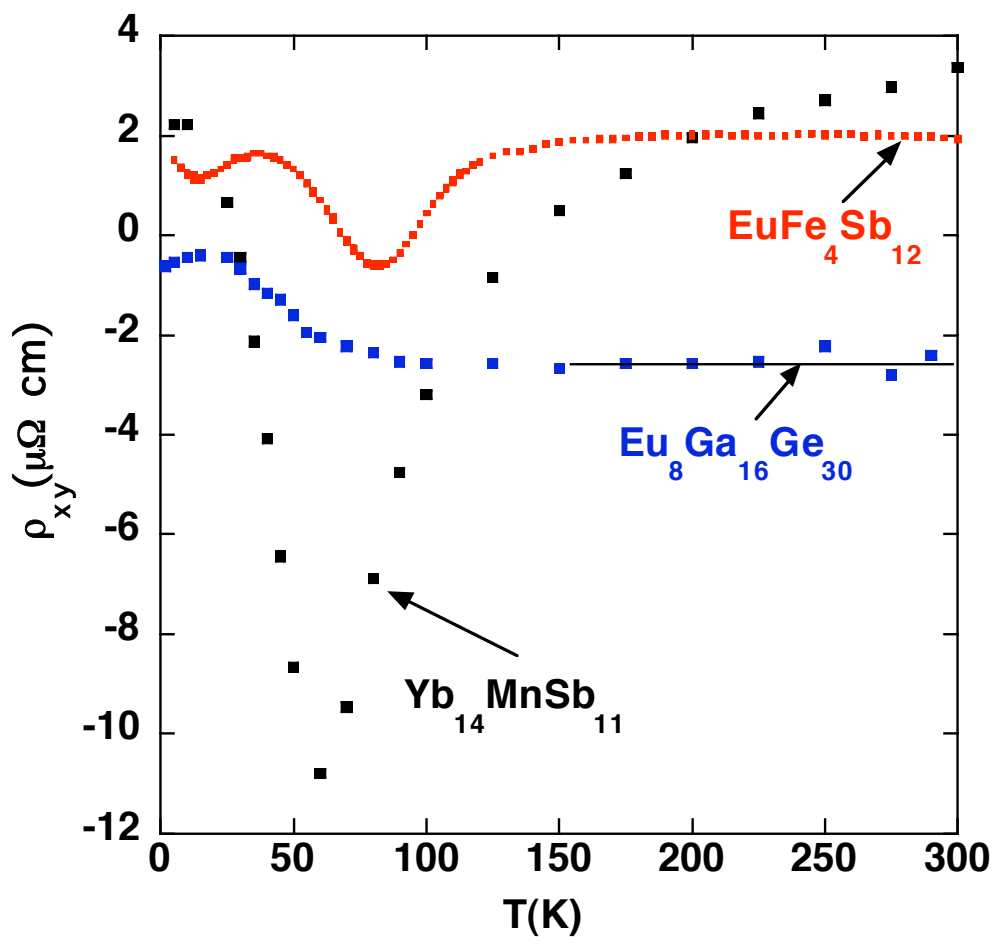

FIG. 5. Hall resistivity vs. temperature data for the three ferromagnets taken in a field of 7 Tesla. For $EuFe_4Sb_{12}$ and $Eu_8Ga_{16}Ge_{30}$ the Hall resistivity is relatively constant for temperatures from 150 to 300 K with values of $2 \times 10^{-6}$ Ω cm and $-2.5 \times 10^{-6}$ Ω cm respectively. Assuming a single carrier band yields a carrier concentration of $2.2 \times 10^{21}$ holes $cm^{-3}$ for $EuFe_4Sb_{12}$ and $1.75 \times 10^{21}$ electrons $cm^{-3}$. The feature near 40 K in the Hall data from $EuFe_4Sb_{12}$ is related to a subtle change in the crystal structure. The Hall resistivity of $Yb_{14}MnSb_{11}$ well above $T_c \approx 53$ K has a significant contribution from the AHE (with an applied field) even at room temperature. Analyses of these data using several models (see text) give a consistent value of $0.93 \times 10^{21}$ holes $cm^{-3}$ for 100 K < T < 300 K.

AHE even at room temperature. This anomalous contribution to the Hall resistivity above $T_c$ (*in an applied field of 7 Tesla*) is large and measurable because at high temperatures the resistivity of $Yb_{14}MnSb_{11}$ is about 3 to 5 times larger than the other two compounds, but also because Kondo physics appears to be important in $Yb_{14}MnSb_{11}$.[19, 28]

As discussed in Ref 18 the Hall data of $Yb_{14}MnSb_{11}$ from 100 K to 300 K can be accurately described by a constant plus a Curie -Weiss term or a constant plus a Curie-Weiss term times the resistivity (or resistivity squared). All three fits to the data give very similar values for the constant, which corresponds to a carrier concentration for this crystal of 9.3 x $10^{20}$ holes $cm^{-3}$. At 5 K the magnetization for each compound is nearly constant for fields greater than 2 T, and the linear slope of the Hall resistivity for H>2T can be used to estimate the carrier concentration (assuming one carrier band). At 5K the carrier concentrations are found to be 3.5 x $10^{21}$ holes $cm^{-3}$, 1.55 x $10^{21}$ electrons $cm^{-3}$, and 1.1 x $10^{21}$ holes $cm^{-3}$, for $EuFe_4Sb_{12}$, $Eu_8Ga_{16}Ge_{30}$, and $Yb_{14}MnSb_{11}$, respectively. These values are not too far from those estimated from the data well above $T_c$.

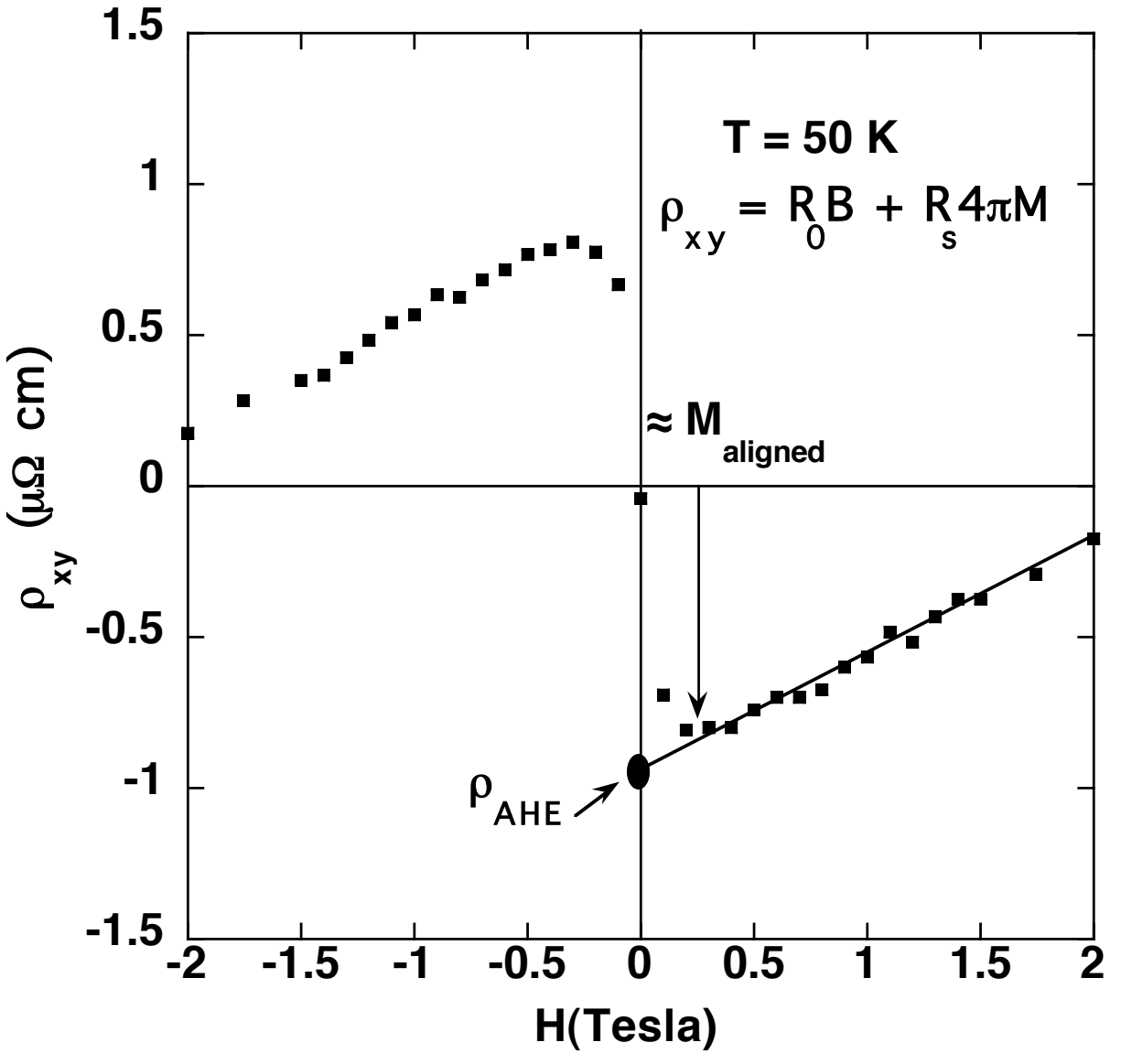

FIG. 6. Example of how the zero field contribution to the anomalous Hall resistivity, $\rho_{AHE} = \rho_{xy}'$ is determined. $M_{aligned}$ is the approximate magnetization where all of the magnetic domains are aligned. The data shown is the measured Hall resistivity of $EuFe_4Sb_{12}$ at 50 K as a function of applied field.

The anomalous Hall resistivity in the limit of zero applied magnetic field is determined from isothermal plots of the Hall resistivity versus applied magnetic field for $T<T_c$. An example of the analysis used is shown in Fig. 6. The Hall resistivity data for fields larger than the value needed to align the domains (denoted as $M_{aligned}$ in the figure) are extrapolated back to H=0. This intercept is the anomalous portion of the Hall resistivity $\rho_{AHE} = \rho_{xy}'$.

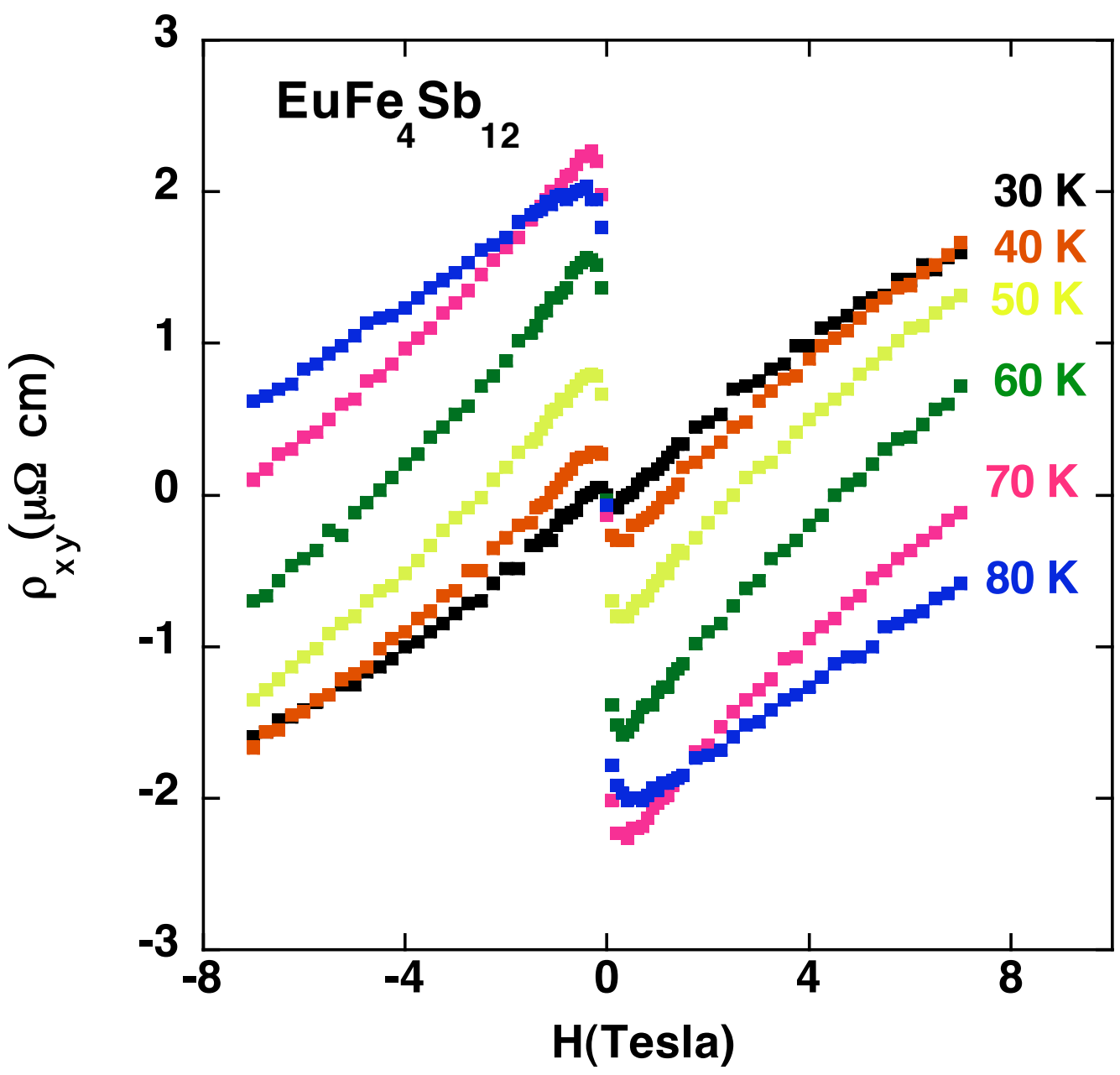

Fig. 7. Isothermal Hall resistivity vs, magnetic field from a polycrystalline sample of $EuFe_4Sb_{12}$.

For each compound, $\rho_{xy}'$ is determined in this manner at several temperatures below $T_c$. The isothermal plots of the Hall resistivity versus field for the three ferromagnets are shown in Figures 7-9.

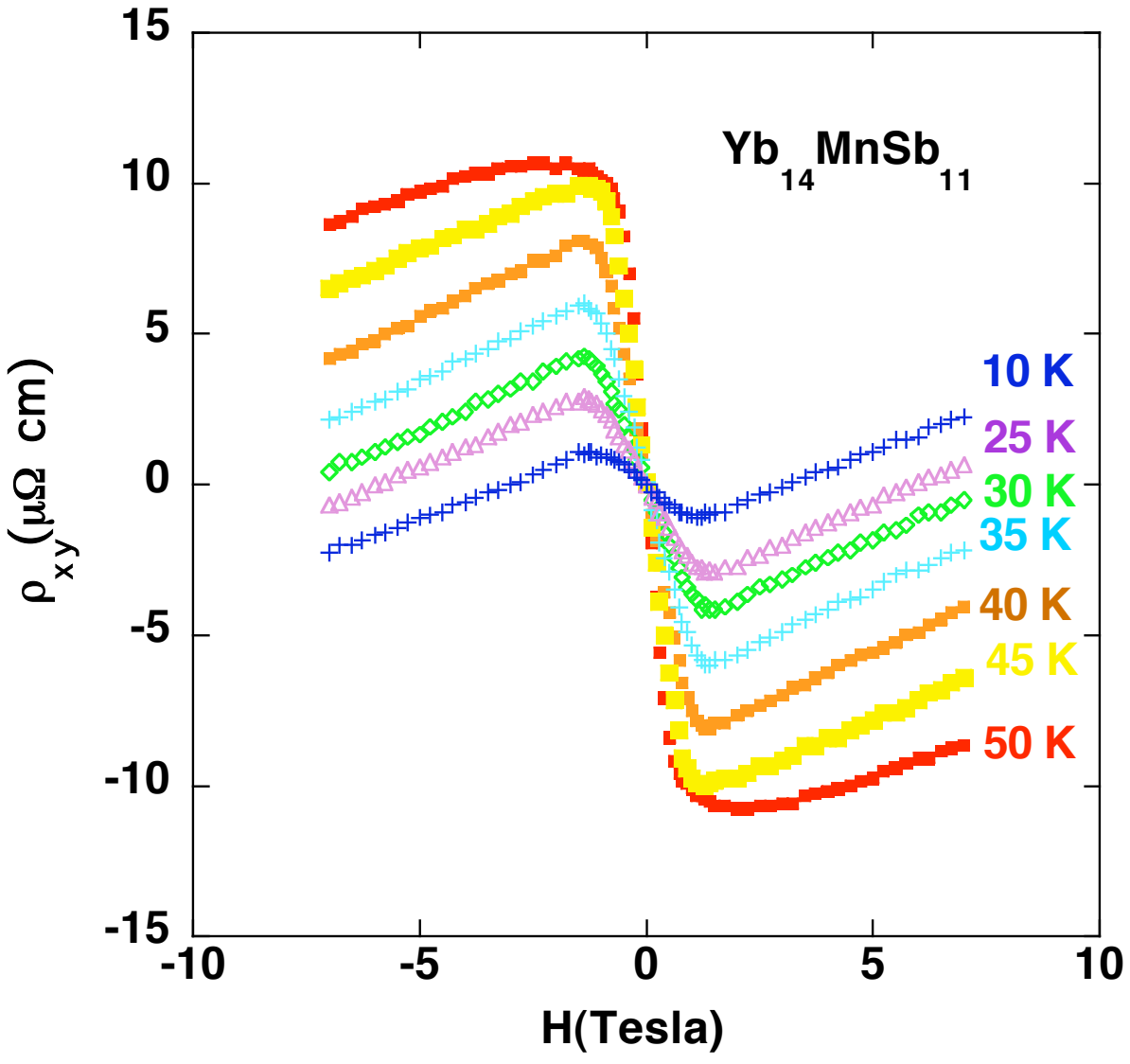

FIG. 8. Isothermal Hall resistivity vs. magnetic field data from an $Yb_{14}MnSb_{11}$ single crystal with H//(110) and current along **c**.

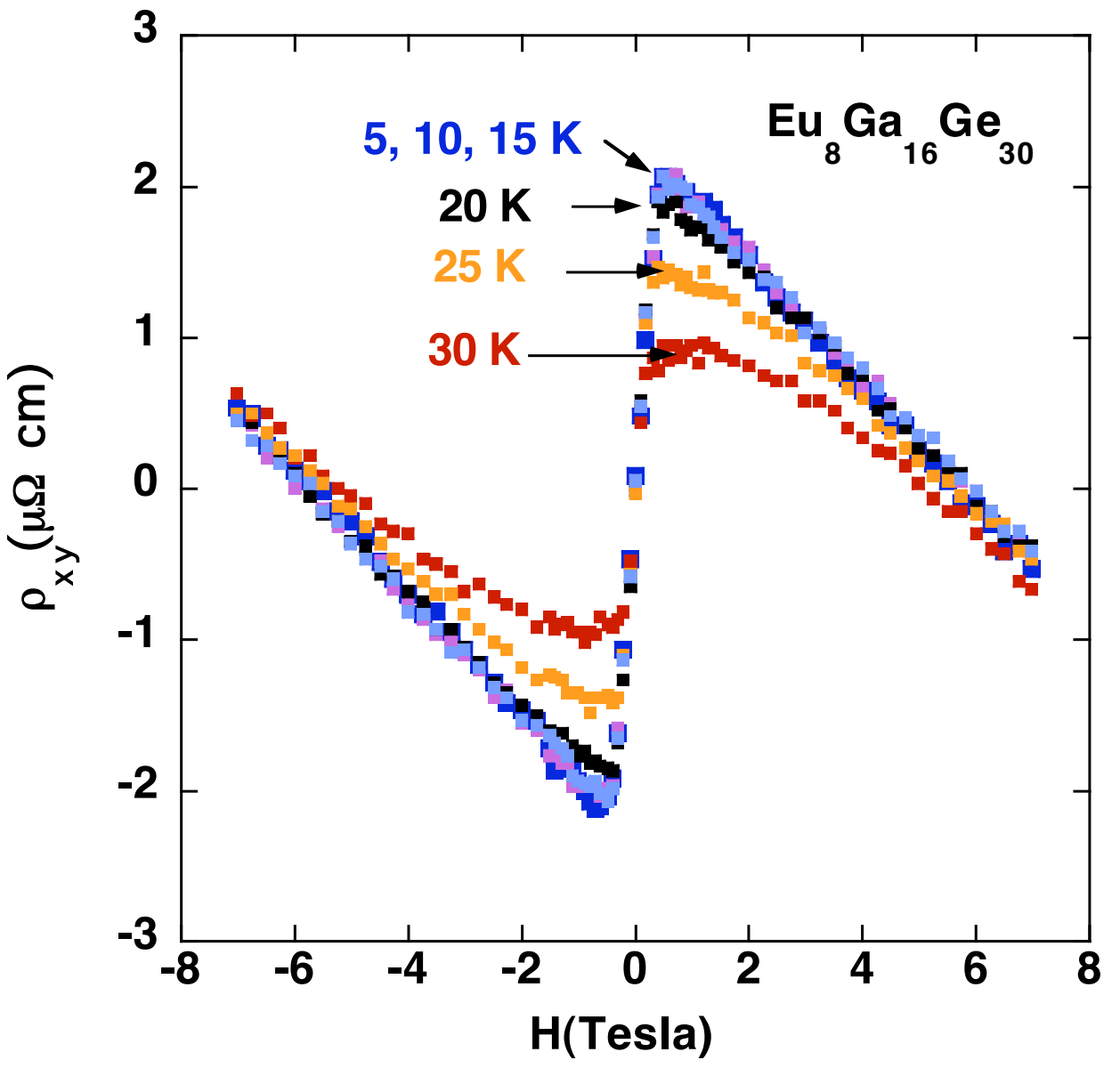

FIG. 9. Isothermal Hall resistivity vs. magnetic field data from an unoriented $Eu_8Ga_{16}Ge_{30}$ single crystal.

## Analysis and Discussion

As discussed in the introduction, theory suggests that $\rho_{xy}'$ can be described by an extrinsic contribution that is proportional to $\rho$ and an intrinsic contribution proportional to $\rho^2$ so that $\rho_{xy}' = \sigma_{xy}' \rho^2 + a\rho$, and to be consistent with recent theoretical calculations $\sigma_{xy}' = \sigma_{xy}$. At low temperatures, $T < 0.2T_c$ , where the spontaneous magnetization is nearly constant, linear fits to plots of $\rho_{xy}'/\rho$ versus $\rho$ can be used to determine $\sigma_{xy}$ and $a$. However, this does not use most of the experimental data that is taken between 0.2 $T_c$ and $\approx$ 0.9 $T_c$. If we assume that $\sigma_{xy}$ and $a$ have no intrinsic variation with temperature except through the variation of the magnetization order parameter with temperature, then $\rho_{xy}' = (M(T)/M(0)) [\sigma_{xy} \rho^2 + a\rho]$. Based on the limited calculations[2] of $\sigma_{xy}$ as a function of temperature, this appears to be a reasonable approximation.

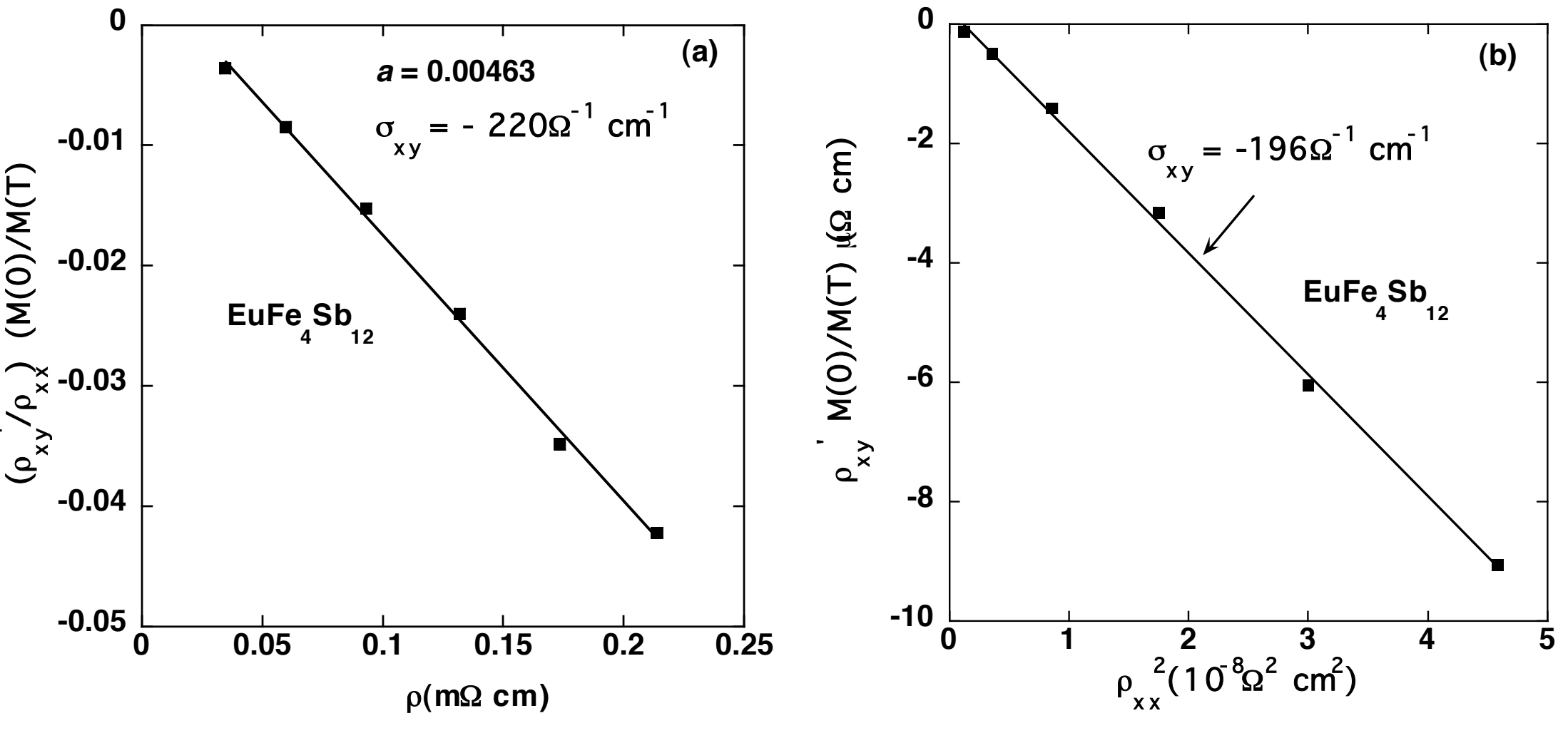

FIG. 10 . (a) Analysis of the intrinsic and extrinsic contributions to the AHE of $EuFe_4Sb_{12}$ from $\rho_{xy}' = (M(T)/M(0)) [\sigma_{xy} \rho^2 + a\rho]$ . (b)Analysis of the intrinsic

contribution to the AHE of $EuFe_4Sb_{12}$ assuming that skew scattering can be neglected ($\rho_{xy}' \approx (M(T)/M(0))\, \sigma_{xy}\, \rho^2$ ).

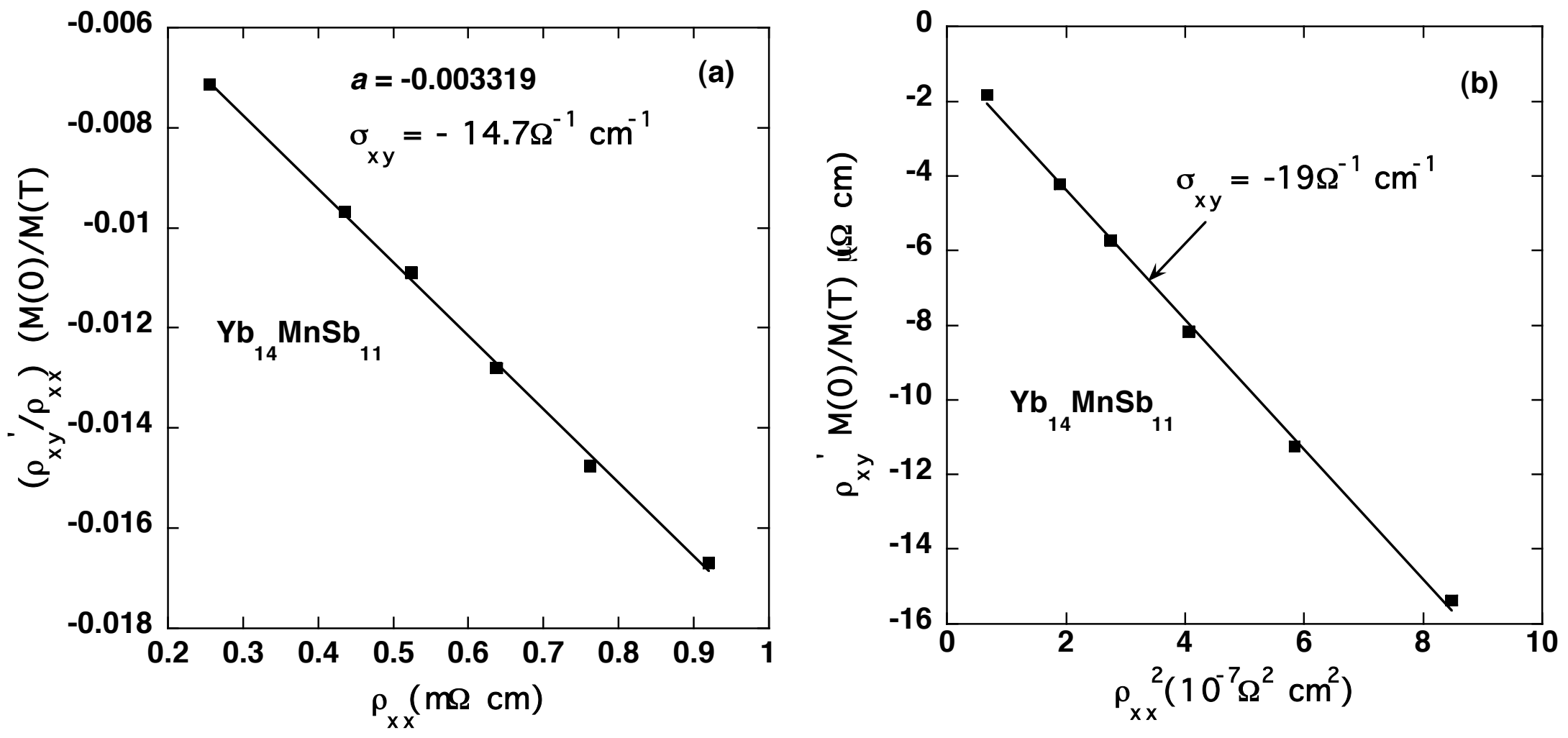

FIG. 11. (a) Analysis of the intrinsic and extrinsic contributions to the AHE of $Yb_{14}MnSb_{11}$ assuming $\rho_{xy}' = (M(T)/M(0))\, [\sigma_{xy}\, \rho^2 + a\rho]$. (b) Analysis of the intrinsic contribution to the AHE of $Yb_{14}MnSb_{11}$ assuming that skew scattering can be neglected ($\rho_{xy}' \approx (M(T)/M(0))\, \sigma_{xy}\, \rho^2$ ).

Plots of M(0)/M(T) [ $\rho_{xy}'/\rho$ ] versus ρ are shown in Fig 10a for $EuFe_4Sb_{12}$. The variation of the magnetization and resistivity with temperature are taken from the data shown in Figs. 3-5. The data are well described by a line with $\sigma_{xy}$ = -220 $\Omega^{-1}$ $cm^{-1}$ and $a$ = 0.0046. If we assume that the extrinsic contribution to $\rho_{xy}'$ from a stochiometric compound can be neglected, i.e. M(0)/M(T) ($\rho_{xy}'$) $\approx \sigma_{xy}\, \rho^2$, then analysis of the $EuFe_4Sb_{12}$ data neglecting skew scattering (Fig. 10b) gives $\sigma_{xy}$ = -196 $\Omega^{-1}$ $cm^{-1}$. Although $EuFe_4Sb_{12}$ exhibits the largest relative decrease in resistivity below $T_c$, the magnitude of $\rho_{xy}'$ below 25 K is below our detectability limit for a bulk sample (see Fig 7). The resistivity of $EuFe_4Sb_{12}$ at T = 2 K is about 7.5 μΩ cm which is too metallic to directly measure the expected value of $\rho_{xy}'$ of 3.5 x $10^{-8}$ Ω cm. A similar analysis of the $Yb_{14}MnSb_{11}$ data shown in Fig 8 results in values of $a$ = -0.0033 and $\sigma_{xy}$ = -14.7 $\Omega^{-1}$ $cm^{-1}$ (Fig 11a) or $\sigma_{xy}$ = -19 $\Omega^{-1}$ $cm^{-1}$ if skew scattering is neglected (Fig 11b). For the $Yb_{14}MnSb_{11}$ crystal the resistivity at our lowest temperature is about 200 μΩ cm, which means that $\rho_{xy}'$ can be directly measured for all temperatures below $T_c$. The resistivity of $Eu_8Ga_{16}Ge_{30}$ only varies by about 30 % below $T_c$ = 36 K . This implies that there is not much variation $\rho_{xy}'$ which makes it impossible to accurately decompose $\rho_{xy}'$ into intrinsic and extrinsic components. Analysis of the data in Fig. 9 assuming M(0)/M(T) $(\rho_{xy}') \approx \sigma_{xy}\,\rho^2$ , however, gives values of $\sigma_{xy} \approx$ 28 ± 3 $\Omega^{-1}$ $cm^{-1}$ (Fig 12). The values of $a$ and $\rho_{xy}$ determined for $EuFe_eSb_{12}$ and $Yb_{14}MnSb_{11}$ are used to estimate the fraction of $\rho_{xy}'$ that is intrinsic as the temperature is varied from T = 0 to about 0.9 $T_c$ (Fig. 13). At the lowest temperatures more than half of $\rho_{xy}'$ is due to an extrinsic contribution linear in ρ such as skew scattering. In spite of this, $\rho_{xy}'$ can also be described fairly well by a single term proportional to $\rho^2$ (Figs. 10b, 11b). This means that it takes a fairly substantial amount of skew scattering at all temperatures before significant deviations from linearity will be observed in plots of M(0)/M(T) $(\rho_{xy}')$ $\approx \sigma_{xy}\,\rho^2$ .

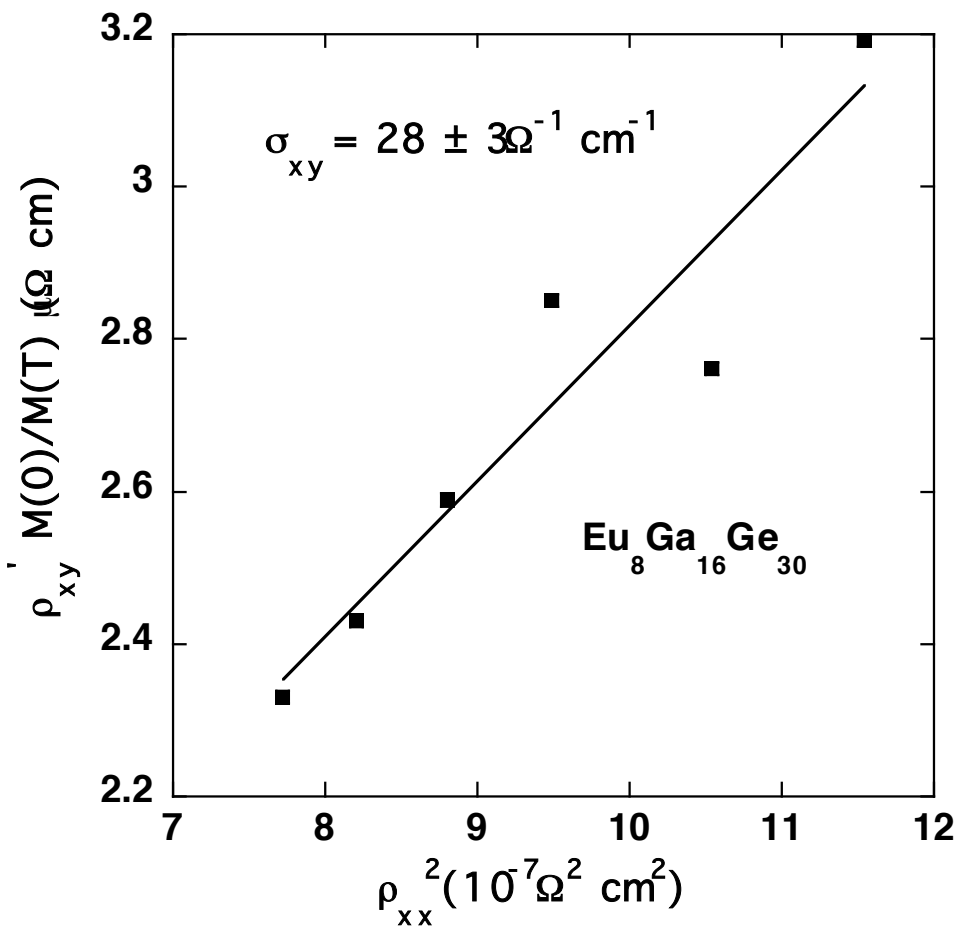

FIG. 12. Approximate analysis of the total anomalous Hall resistivity vs temperature of $Eu_8Ga_{16}Ge_{30}$ assuming $\rho_{xy}' \approx (M(T)/M(0))\,\sigma_{xy}\,\rho^2$ . The small variation of the resistivity of $Eu_8Ga_{16}Ge_{30}$ below $T_c$ makes it impossible to accurately separate the Hall resistivity into intrinsic and extrinsic components.

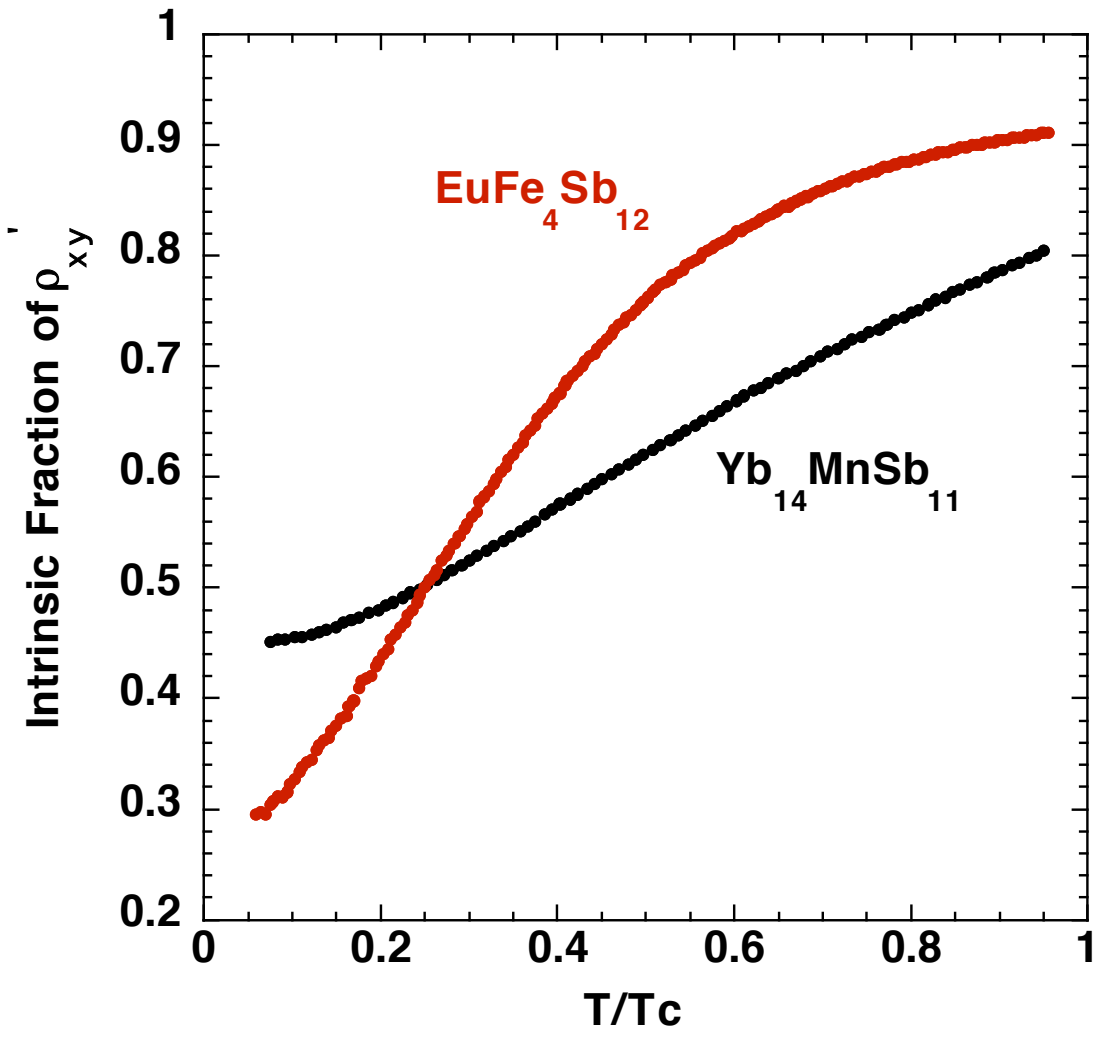

FIG. 13. Variation of the intrinsic fraction of the AHE with temperature as determined from the fits to the data in Figs 10a, and 11a, and using the resistivity data in Fig 4.

Some relevant properties from all three ferromagnets are summarized in Table 1.There are several important observations. There is clearly a substantial intrinsic contribution to $\rho_{xy}'$ as T approaches 0. This result supports recent theories,[1,2,3] which attribute the intrinsic AHE to a ground state property that can be determined from electronic structure calculations. The excellent linear variation of M(0)/M(T) [ $\rho_{xy}'/\rho$ ] versus $\rho$ over a wide range of temperatures and resistivities implies that both *a* and $\sigma_{xy}$ can only weakly depend on temperature. A weak variation of $\sigma_{xy}$ with temperature was theoretically predicted for iron at temperatures below 300 K. The present results suggest that this weak variation may hold for all temperatures below $T_c$ if the temperature variation of the spontaneous magnetization is taken into account. For all three ferromagnets, the sign of $\sigma_{xy}$ is opposite to the normal Hall component. This was also the case in a careful recent study on Co films[12] but in older studies[35] on Gd crystals,and on various transition metal films[36] $\sigma_{xy}$ appears to have the same sign as the normal Hall component. From our understanding of current theory[1,2] on the AHE, there does not seem to be a simple rule as to the sign of $\sigma_{xy}$ relative to the sign of the carriers. The magnitude of $\sigma_{xy} = 220\ \Omega^{-1}cm^{-1}$ for $EuFe_4Sb_{12}$ is much larger than the values of $\sigma_{xy} \approx 10$ to30 $\Omega^{-1}cm^{-1}$ found for the other two ferromagnets. The value for $EuFe_4Sb_{12}$ is similar to the value of 240 $\Omega^{-1}cm^{-1}$ found for Co films and is probably related to the weak itinerant magnetism associated with the Fe 3d bands. It would be interesting if theory could calculate $\sigma_{xy}$ for $EuFe_4Sb_{12}$ or the related weak ferromagnet $NaFe_4Sb_{12}$.[31] While the values of $\sigma_{xy}$ for $Yb_{14}MnSb_{11}$ and $Eu_8Ga_{16}Ge_{30}$ are substantially smaller than found for $EuFe_4Sb_{12}$, they are comparable to the "gigantic anomalous Hall effect" found[5] in the magnetically complex pyrochlore $Nd_2Mo_2O_7$. Another curious observation is the magnitude of the extrinsic skew scattering coefficient. *a,* is about the same for the two crystals where it could be accurately determined ( $|a| \approx 0.004$). This may indicate a similar degree of disorder and defects in the two materials, but we really do not know. Finally, the recent ideas of the AHE have only been carefully tested on relatively simple materials such as Fe[2] and Co[12] and very recently $Mn_5Ge_3$[37]. The apparent applicability of these ideas to considerably more complex ferromagnets, bodes well for our general understanding of the origin of the AHE.

**Table 1.**

**Selected properties of the three ferromagnets**

| Compound | $EuFe_4Sb_{12}$ | $Yb_{14}MnSb_{11}$ | $Eu_8Ga_{16}Ge_{30}$ |
|---|---|---|---|
| $T_c$(K) | 84 | 53 | 36 |
| $4\pi M_o$(gauss) | 1560 | 615 | 5535 |
| $n_{high\ T}$ ($10^{21}$ $cm^{-3}$) | 2.2 | 0.93 | -1.75 |
| $n_{5K}$ ($10^{21}$ $cm^{-3}$) | 3.54 | 1.1 | -1.55 |
| *a* | 0.0046 | -0.0033 | ____ |
| $\sigma_{xy}$ ( $\Omega^{-1}cm^{-1}$) | -220 | -14.7 | 28 ± 3 |

## Acknowledgements

It is a pleasure to acknowledge stimulating and illuminating discussions with Alan MacDonald, David Singh,Thomas Schulthess, V. V. “Krishna” Krishnamurthy, Veerle Keppens, Sriparna Bhattacharya and Raphael Hermann. Research sponsored by the Division of Materials Sciences and Engineering,Office of Basic Energy Sciences, U.S. Department of Energy, under contract DE-AC05-00OR22725 with Oak Ridge NationalLaboratory, managed and operated by UT-Battelle, LLC.